\documentclass[aps,prc,twocolumn,showpacs,floatfix,nofootinbib]{revtex4}
\usepackage{amsmath,graphicx}
\usepackage{float}

\begin{document}

\title{Relativistic third-order dissipative fluid dynamics from kinetic theory}

\author{Amaresh Jaiswal}
\affiliation{Tata Institute of Fundamental Research,
Homi Bhabha Road, Mumbai 400005, India}

\date{\today}

\begin{abstract}
We present the derivation of a novel third-order hydrodynamic 
evolution equation for shear stress tensor from kinetic theory. 
Boltzmann equation with relaxation time approximation for the 
collision term is solved iteratively using Chapman-Enskog like 
expansion to obtain the nonequilibrium phase-space distribution 
function. Subsequently, the evolution equation for shear stress 
tensor is derived from its kinetic definition up-to third-order in 
gradients. We quantify the significance of the new derivation within 
one-dimensional scaling expansion and demonstrate that the results 
obtained using third-order viscous equations derived here provides a 
very good approximation to the exact solution of Boltzmann equation 
in relaxation time approximation. We also show that the time 
evolution of pressure anisotropy obtained using our equations is in 
better agreement with transport results when compared with an 
existing third-order calculation based on the second-law of 
thermodynamics.
\end{abstract}

\pacs{05.20.Dd, 47.75+f, 47.10.-g, 47.10.A-}


\maketitle


Fluid dynamics is an effective theory describing the long-wavelength 
limit of the microscopic dynamics of a system. While nonrelativistic 
fluid dynamics finds applications in various aspects of physics and 
engineering, the domain of applicability of relativistic fluid 
dynamics is in the field of astrophysics, cosmology and high-energy 
heavy-ion collisions. The collective behavior of the hot and dense 
matter (which is believed to have existed in the very early 
universe) created in ultra-relativistic heavy-ion collisions has 
been studied quite extensively within the framework of relativistic 
fluid dynamics. To study the evolution of a hydrodynamic system, it 
is natural to first employ the equations of ideal fluid dynamics. 
However, as ideal fluids are hypothetical by virtue of uncertainty 
principle \cite {Danielewicz:1984ww}, the dissipative effects can 
not be ignored.

Relativistic dissipative fluid dynamics is formulated as an 
order-by-order expansion in gradients, ideal hydrodynamics being 
zeroth-order. The first-order theories, collectively known as 
relativistic Navier-Stokes (NS) theory, are due to Eckart \cite 
{Eckart:1940zz} and Landau-Lifshitz \cite{Landau}. However, these 
theories involve parabolic differential equations and suffer from 
acausality and numerical instability. The second-order theory by 
Israel and Stewart (IS) \cite {Israel:1979wp} with its hyperbolic 
equations solves the acausality problem \cite {Huovinen:2008te} but 
may not guarantee stability. Despite the success of IS theory in 
explaining a wide range of collective phenomena observed in 
heavy-ion collisions, its formulation is based on strong assumptions 
and approximations. The original IS theory derived from Boltzmann 
equation (BE) uses two powerful assumptions in the derivation of 
dissipative equations: use of second moment of BE and the 14-moment 
approximation \cite{Israel:1979wp, Grad}. In Ref. \cite 
{Denicol:2010xn}, although the dissipative equations were derived 
directly from their definitions without resorting to second-moment 
of BE, however the 14-moment approximation was still employed. In 
Ref. \cite {Jaiswal:2013npa} it was shown that both these 
assumptions are unnecessary and instead of 14-moment approximation, 
iterative solution of BE was used to obtain the dissipative 
evolution equations from their definitions.

Apart from these problems in the formulation, IS theory suffers from 
several other shortcomings. In one-dimensional Bjorken scaling 
expansion \cite {Bjorken:1982qr}, for large viscosities or small 
initial time, IS theory has resulted in unphysical effects such as 
reheating of the expanding medium \cite {Muronga:2003ta} and 
negative longitudinal pressure \cite {Martinez:2009mf}. Further, the 
scaling solutions of IS equations when compared with transport 
results show disagreement for $\eta/s>0.5$ indicating the breakdown 
of second-order theory \cite {Huovinen:2008te,El:2008yy}. With this 
motivation, in Ref. \cite {Jaiswal:2012qm}, second-order dissipative 
equations were derived from BE where the collision term was 
generalized to include nonlocal effects via gradients of the 
distribution function. Moreover, in Refs. \cite 
{Jaiswal:2013npa,El:2009vj} it was demonstrated that a heuristic 
inclusion of higher-order corrections led to an improved agreement 
with transport results. In fact, the derivation of higher-order 
constitutive equations from kinetic theory for non-relativistic 
systems has been known for a long time \cite{Cha}. Thus it is of 
interest to improvise the relativistic second-order theory by 
incorporating higher-order corrections. 

In this Rapid Communication, we derive a new relativistic 
third-order evolution equation for shear stress tensor from kinetic 
theory. Without resorting to the widely used Grad's 14-moment 
approximation \cite{Grad}, we iteratively solve the BE in relaxation 
time approximation (RTA) to obtain nonequilibrium phase-space 
distribution function. We subsequently derive equation of motion for 
shear stress tensor up-to third-order, directly from its definition. 
Within one-dimensional scaling expansion, the results obtained using 
third-order evolution equations derived here shows improved 
agreement with exact solution of BE as compared to second-order 
equations. We also demonstrate that the evolution of pressure 
anisotropy obtained using our equations shows better agreement with 
the transport results as compared to those obtained by using an 
existing third-order equation derived from entropy considerations.


The hydrodynamic evolution of a system is governed by the 
conservation equations for energy and momentum. The conserved 
energy-momentum tensor can be expressed in terms of single-particle, 
phase-space distribution function and tensor decomposed into 
hydrodynamic variables \cite{deGroot}. For a system of massless 
particles, bulk viscosity vanishes leading to
\begin{align}\label{NTD}
T^{\mu\nu} &= \!\int\! dp \ p^\mu p^\nu\, f(x,p) = \epsilon u^\mu u^\nu 
- P\Delta ^{\mu \nu} + \pi^{\mu\nu},
\end{align}
where $dp\equiv g d{\bf p}/[(2 \pi)^3|\bf p|]$, $g$ being the 
degeneracy factor, $p^\mu$ is the particle four-momentum and $f(x,p)$
is the phase-space distribution function. In the tensor 
decompositions, $\epsilon$, $P$ and $\pi^{\mu\nu}$ are respectively 
energy density, pressure and the shear stress tensor. The projection 
operator $\Delta^{\mu\nu}\equiv g^{\mu\nu}-u^\mu u^\nu$ is 
orthogonal to the hydrodynamic four-velocity $u^\mu$ defined in the 
Landau frame: $T^{\mu\nu} u_\nu=\epsilon u^\mu$. The metric tensor 
is Minkowskian, $g^{\mu\nu}\equiv\mathrm{diag}(+,-,-,-)$.

Energy-momentum conservation, $\partial_\mu T^{\mu\nu} =0$ yields 
the fundamental evolution equations for $\epsilon$ and $u^\mu$
\begin{align}\label{evol}
\dot\epsilon + (\epsilon+P)\theta - \pi^{\mu\nu}\nabla_{(\mu} u_{\nu)} &= 0,  \nonumber\\
(\epsilon+P)\dot u^\alpha - \nabla^\alpha P + \Delta^\alpha_\nu \partial_\mu \pi^{\mu\nu}  &= 0. 
\end{align}
We use the notation $\dot A\equiv u^\mu\partial_\mu A$ for comoving 
derivative, $\theta\equiv \partial_\mu u^\mu$ for the expansion 
scalar, $A^{(\alpha}B^{\beta )}\equiv (A^\alpha B^\beta + A^\beta 
B^\alpha)/2$ for symmetrization and $\nabla^\alpha\equiv 
\Delta^{\mu\alpha}\partial_\mu$ for space-like derivative. In the 
massless limit, the energy density and pressure are related as 
$\epsilon=3P\propto\beta^{-4}$. The inverse temperature, 
$\beta\equiv1/T$, is defined by the Landau matching condition 
$\epsilon=\epsilon_0$ where $\epsilon_0$ is the equilibrium energy 
density. In this limit, Eqs. (\ref{evol}) can be used to obtain the 
derivatives of $\beta$ as
\begin{equation}\label{evol1}
\dot\beta = \frac{\beta}{3}\theta - \frac{\beta}{12P}\pi^{\rho\gamma}\sigma_{\rho\gamma}, ~~
\nabla^\alpha\beta = \!-\beta\dot u^\alpha - \frac{\beta}{4P} \Delta^\alpha_\rho \partial_\gamma \pi^{\rho\gamma} , 
\end{equation}
where 
$\sigma^{\rho\gamma}\equiv\nabla^{(\rho}u^{\gamma)}-(\theta/3)\Delta^{\rho
\gamma}$ is the velocity stress tensor. The above identities will 
be helpful in the derivation of shear evolution equation.

The expression for shear stress tensor ($\pi^{\mu\nu}$) can be 
obtained in terms of the out-of-equilibrium part of the distribution 
function. To this end, we write the nonequilibrium distribution 
function as $f=f_0+\delta f$, where the deviation from equilibrium 
is assumed to be small $(\delta f\ll f)$. The equilibrium 
distribution function represents Boltzmann statistics of massless 
particles at vanishing chemical potential, $f_0=\exp(-\beta\,u\cdot 
p)$, where $u\cdot p \equiv u_\mu p^\mu$. From Eq. (\ref{NTD}), 
$\pi^{\mu\nu}$ can be expressed in terms of $\delta f$ as
\begin{align}\label{FSE}
\pi^{\mu\nu} &= \Delta^{\mu\nu}_{\alpha\beta} \int dp \, p^\alpha p^\beta\, \delta f,
\end{align}
where $\Delta^{\mu\nu}_{\alpha\beta}\equiv 
\Delta^{\mu}_{(\alpha}\Delta^{\nu}_{\beta)} - 
(1/3)\Delta^{\mu\nu}\Delta_{\alpha\beta}$ is a traceless symmetric 
projection operator orthogonal to $u^\mu$. To proceed further, the 
form of $\delta f$ has to be specified. In the following, Boltzmann 
equation in RTA will be solved iteratively to obtain $\delta f$ 
order-by-order in gradients.
 

Nonequilibrium phase-space distribution function can be obtained by 
solving the one-body kinetic equation such as the Boltzmann 
equation. The most common technique of generating solutions to such 
equations is the Chapman-Enskog expansion where the particle 
distribution function is expanded about its equilibrium value in 
powers of space-time gradients \cite{Chapman}
\begin{equation}\label{CEE}
f = f_0 + \delta f, \quad \delta f= \delta f^{(1)} + \delta f^{(2)} + \cdots,
\end{equation}
where $\delta f^{(1)}$ is first-order in derivatives, $\delta f^{(2)}$ 
is second-order and so on. Subsequently, the relativistic Boltzmann 
equation with relaxation time approximation for the collision term 
\cite{Anderson_Witting},
\begin{equation}\label{RBE}
p^\mu\partial_\mu f =  -u\!\cdot\! p\frac{\delta f}{\tau_R} \;\Rightarrow\; 
f=f_0-(\tau_R/u\!\cdot\! p)\,p^\mu\partial_\mu f,
\end{equation}
can be solved iteratively as \cite{Jaiswal:2013npa,Romatschke:2011qp}
\begin{equation}\label{F1F2}
f_1 = f_0 -\frac{\tau_R}{u\!\cdot\! p} \, p^\mu \partial_\mu f_0, \quad 
f_2 = f_0 -\frac{\tau_R}{u\!\cdot\! p} \, p^\mu \partial_\mu f_1, ~~\, \cdots
\end{equation}
where $f_n=f_0+\delta f^{(1)}+\delta f^{(2)}+\cdots+\delta f^{(n)}$. 
To first and second-order in derivatives, we obtain
\begin{align}
\delta f^{(1)} &= -\frac{\tau_R}{u\!\cdot\! p} \, p^\mu \partial_\mu f_0, \label{FOC} \\
\delta f^{(2)} &= \frac{\tau_R}{u\!\cdot\! p}p^\mu p^\nu\partial_\mu\Big(\frac{\tau_R}{u\!\cdot\! p} \partial_\nu f_0\Big). \label{SOC}
\end{align}
The above expressions for nonequilibrium part of the distribution 
function along with Eq. (\ref{FSE}) will be used in the derivation 
of shear evolution equations.

As a side remark, note that the RTA for the collision term, 
$C[f]=-(u\cdot p)\delta f/\tau_R$ in Eq. (\ref {RBE}), should 
satisfy current and energy-momentum conservation, i.e., the zeroth 
and first moment of the collision term should vanish \cite 
{deGroot}. Assuming the relaxation time $\tau_R$ to be independent 
of momenta, these conservation equations are satisfied only if the 
fluid four-velocity is defined in the Landau frame \cite 
{Anderson_Witting}. Hence, within RTA, the Landau frame is imposed 
and is not a choice.


The first-order expression for shear stress tensor can be obtained 
from Eq. (\ref{FSE}) using $\delta f = \delta f^{(1)}$ from Eq. (\ref
{FOC}),
\begin{align}
\pi^{\mu\nu} &= \Delta^{\mu\nu}_{\alpha\beta}\int dp \ p^\alpha p^\beta \left(-\frac{\tau_R}{u\!\cdot\! p} \, p^\mu \partial_\mu\, f_0\right) . \label{FOSE}
\end{align}
Using Eqs. (\ref{evol1}) and keeping only those terms which are 
first-order in gradients, the integrals in the above equation reduce 
to
\begin{equation}\label{FOE}
\pi^{\mu\nu} = 2\tau_R\beta_\pi\sigma^{\mu\nu}, \quad \beta_\pi = \frac{4}{5}P .
\end{equation}

To obtain the second-order evolution equation, we follow the 
methodology discussed in Ref. \cite {Denicol:2010xn}. The evolution 
of the shear stress tensor can be obtained by considering the 
comoving derivative of Eq. (\ref{FSE}),
\begin{equation}
\dot\pi^{\langle\mu\nu\rangle} = \Delta^{\mu\nu}_{\alpha\beta} \int dp\, p^\alpha p^\beta\, \delta\dot f, \label{SSE}
\end{equation}
where the notation $A^{\langle\mu\nu\rangle}\equiv 
\Delta^{\mu\nu}_{\alpha\beta}A^{\alpha\beta}$ represents traceless 
symmetric projection orthogonal to $u^{\mu}$.

The comoving derivative of the nonequilibrium part of the 
distribution function ($\delta\dot f$) can be obtained by rewriting 
Eq. (\ref{RBE}) in the form
\begin{equation}\label{DFD}
\delta\dot f = -\dot f_0 - \frac{1}{u\!\cdot\! p}p^\gamma\nabla_\gamma f - \frac{\delta f}{\tau_R},
\end{equation}
Using this expression for $\delta\dot f$ in Eq. (\ref{SSE}), we obtain
\begin{equation} 
\dot\pi^{\langle\mu\nu\rangle} + \frac{\pi^{\mu\nu}}{\tau_R} = 
- \Delta^{\mu\nu}_{\alpha\beta} \!\int\! dp \, p^\alpha p^\beta \!\left(\dot f_0 + \frac{1}{u\!\cdot\! p}p^\gamma\nabla_\gamma f\right)\!. \label{SOSE}
\end{equation}
It is clear that in the above equation, the Boltzmann relaxation 
time $\tau_R$ can be replaced by the shear relaxation time 
$\tau_\pi$. By comparing the first-order evolution Eq. (\ref {FOE}) 
with the relativistic Navier-Stokes equation 
$\pi^{\mu\nu}=2\eta\sigma^{\mu\nu}$, the shear relaxation time is 
obtained in terms of the first-order transport coefficient, 
$\tau_\pi=\eta/\beta_\pi$.

Note that for the shear evolution equations to be second-order in 
gradients, the distribution function on the right hand side of Eq. 
(\ref {SOSE}) need to be computed only till first-order, i.e., 
$f=f_1=f_0+\delta f^{(1)}$. Using Eq. (\ref{FOC}) for $\delta 
f^{(1)}$ and Eqs. (\ref{evol1}) for derivatives of $\beta$, and 
keeping terms upto quadratic order in gradients, the second-order 
shear evolution equation is obtained as \cite {Jaiswal:2013npa}
\begin{equation}\label{SOSHEAR}
\dot{\pi}^{\langle\mu\nu\rangle} \!+ \frac{\pi^{\mu\nu}}{\tau_\pi}\!= 
2\beta_{\pi}\sigma^{\mu\nu}
\!+2\pi_\gamma^{\langle\mu}\omega^{\nu\rangle\gamma}
\!-\frac{10}{7}\pi_\gamma^{\langle\mu}\sigma^{\nu\rangle\gamma} 
\!-\frac{4}{3}\pi^{\mu\nu}\theta,
\end{equation}
where $\omega^{\mu\nu}\equiv(\nabla^\mu u^\nu-\nabla^\nu u^\mu)/2$ 
is the vorticity tensor. We have used the first-order expression for 
shear stress tensor, Eq. (\ref{FOE}), to replace 
$\sigma^{\mu\nu}\to\pi^{\mu\nu}$ such that the relaxation times 
appearing on the right hand side of Eq. (\ref {SOSE}) are absorbed.

To derive a third-order evolution equation for shear stress tensor, 
the distribution function on the right hand side of Eq. (\ref 
{SOSE}) needs to be computed till second-order ($\delta f=\delta 
f^{(1)}+\delta f^{(2)}$). In order to account for all the 
higher-order terms, Eq. (\ref{SOSHEAR}) was used to substitute for 
$\sigma^{\mu\nu}$. Employing Eqs. (\ref{evol1}) for derivatives of 
$\beta$ and keeping terms upto cubic order in derivatives, we 
finally obtain a unique third-order evolution equation for shear 
stress tensor after a straightforward but tedious algebra
\begin{align}\label{TOSHEAR}
\dot{\pi}^{\langle\mu\nu\rangle} =& -\frac{\pi^{\mu\nu}}{\tau_\pi}
+2\beta_\pi\sigma^{\mu\nu}
+2\pi_{\gamma}^{\langle\mu}\omega^{\nu\rangle\gamma}
-\frac{10}{7}\pi_\gamma^{\langle\mu}\sigma^{\nu\rangle\gamma}  \nonumber \\
&-\frac{4}{3}\pi^{\mu\nu}\theta
+\frac{25}{7\beta_\pi}\pi^{\rho\langle\mu}\omega^{\nu\rangle\gamma}\pi_{\rho\gamma}
-\frac{1}{3\beta_\pi}\pi_\gamma^{\langle\mu}\pi^{\nu\rangle\gamma}\theta \nonumber \\
&-\frac{38}{245\beta_\pi}\pi^{\mu\nu}\pi^{\rho\gamma}\sigma_{\rho\gamma}
-\frac{22}{49\beta_\pi}\pi^{\rho\langle\mu}\pi^{\nu\rangle\gamma}\sigma_{\rho\gamma} \nonumber \\
&-\frac{24}{35}\nabla^{\langle\mu}\left(\pi^{\nu\rangle\gamma}\dot u_\gamma\tau_\pi\right)
+\frac{4}{35}\nabla^{\langle\mu}\left(\tau_\pi\nabla_\gamma\pi^{\nu\rangle\gamma}\right) \nonumber \\
&-\frac{2}{7}\nabla_{\gamma}\left(\tau_\pi\nabla^{\langle\mu}\pi^{\nu\rangle\gamma}\right)
+\frac{12}{7}\nabla_{\gamma}\left(\tau_\pi\dot u^{\langle\mu}\pi^{\nu\rangle\gamma}\right) \nonumber \\
&-\frac{1}{7}\nabla_{\gamma}\left(\tau_\pi\nabla^{\gamma}\pi^{\langle\mu\nu\rangle}\right)
+\frac{6}{7}\nabla_{\gamma}\left(\tau_\pi\dot u^{\gamma}\pi^{\langle\mu\nu\rangle}\right) \nonumber \\
&-\frac{2}{7}\tau_\pi\omega^{\rho\langle\mu}\omega^{\nu\rangle\gamma}\pi_{\rho\gamma}
-\frac{2}{7}\tau_\pi\pi^{\rho\langle\mu}\omega^{\nu\rangle\gamma}\omega_{\rho\gamma} \nonumber \\
&-\frac{10}{63}\tau_\pi\pi^{\mu\nu}\theta^2
+\frac{26}{21}\tau_\pi\pi_\gamma^{\langle\mu}\omega^{\nu\rangle\gamma}\theta.
\end{align}
This is the main result of the present work. We note that Eq. (\ref
{TOSHEAR}) represents only a subset of all possible third order 
terms because bulk viscosity and heat current has been neglected.

We compare the third-order shear evolution equation derived in the 
present work with that obtained in Ref. \cite {El:2009vj}. In the 
latter work, the shear evolution equation was derived by invoking 
second law of thermodynamics from kinetic definition of entropy 
four-current, expanded till third-order in $\pi^{\mu\nu}$. For ease 
of comparison, we write the evolution equation obtained in Ref. \cite
{El:2009vj} in the form
\begin{align}\label{TOEF}
\dot\pi^{\langle\mu\nu\rangle} =& -\frac{\pi^{\mu\nu}}{\tau_\pi'} + 2\beta_\pi'\sigma^{\mu\nu}
-\frac{4}{3}\pi^{\mu\nu}\theta 
+ \frac{5}{36\beta_\pi'}\pi^{\mu\nu}\pi^{\rho\gamma}\sigma_{\rho\gamma} \nonumber \\
&-\frac{16}{9\beta_\pi'}\pi^{<\mu}_{\gamma}\pi^{\nu>\gamma}\theta,
\end{align}
where $\beta_\pi'=2P/3$ and $\tau_\pi'=\eta/\beta_\pi'$. We observe 
that the right-hand-side of Eq. (\ref{TOEF}) contains one 
second-order and two third-order terms compared to three 
second-order and fourteen third-order terms obtained in the present 
work, i.e.,  Eq. (\ref{TOSHEAR}). It is well known that the approach 
based on entropy method fails to capture all the terms in the 
dissipative evolution equations even at second-order. Moreover, the 
discrepancy at third-order confirms the fact that the evolution 
equation obtained by invoking second law of thermodynamics is 
incomplete.


To demonstrate the numerical significance of the third-order shear 
evolution equation derived here, we consider boost-invariant Bjorken 
expansion of a system consisting of massless Boltzmann gas \cite 
{Bjorken:1982qr}. Working in Milne coordinates $(\tau,x,y,\eta_s)$, 
where $\tau = \sqrt{t^2-z^2}$, $\eta_s=\tanh^{-1}(z/t)$, and with 
$u^\mu=(1,0,0,0)$, we observe that only the $\eta_s\eta_s$ component 
of Eq. (\ref{TOSHEAR}) survives. In this scenario, 
$\omega^{\mu\nu}=\dot u^\mu=\nabla^\mu\tau_\pi=0$, $\theta = 1/\tau$ 
and $\sigma^{\eta_s\eta_s} = -2/(3\tau^3)$. Defining 
$\pi\equiv-\tau^2\pi^{\eta_s\eta_s}$, we find that
$\pi^{\rho\gamma}\sigma_{\rho\gamma} = \pi/\tau$, and
\begin{align}
&\dot\pi^{\langle\eta_s\eta_s\rangle} \!=\! -\frac{1}{\tau^2}\frac{d\pi}{d\tau}, ~
\pi^{\langle\eta_s}_{\gamma}\sigma^{\eta_s\rangle\gamma} \!=\! -\frac{\pi}{3\tau^3}, ~
\pi^{\langle\eta_s}_{\gamma}\pi^{\eta_s\rangle\gamma} \!=\! -\frac{\pi^2}{2\tau^2}, \nonumber\\ 
&\pi^{\rho\langle\eta_s}\pi^{\eta_s\rangle\gamma}\sigma_{\rho\gamma} = -\frac{\pi^2}{2\tau^3}, \quad
\nabla^{\langle\eta_s}\nabla_{\gamma}\pi^{\eta_s\rangle\gamma} = \frac{2\pi}{3\tau^4}, \nonumber\\ 
&\nabla_{\gamma}\nabla^{\langle\eta_s}\pi^{\eta_s\rangle\gamma} = \frac{4\pi}{3\tau^4}, \quad
\nabla^2\pi^{\langle\eta_s\eta_s\rangle} = \frac{4\pi}{3\tau^4}.
\label{identity}
\end{align} 
Using the above results, evolution of $\epsilon$ and $\pi$ from Eqs. 
(\ref{evol}) and (\ref{TOSHEAR}) reduces to
\begin{align}
\frac{d\epsilon}{d\tau} &= -\frac{1}{\tau}\left(\epsilon + P  -\pi\right), \label{BED} \\
\frac{d\pi}{d\tau} &= - \frac{\pi}{\tau_\pi} + \beta_\pi\frac{4}{3\tau} - \lambda\frac{\pi}{\tau} - \chi\frac{\pi^2}{\beta_\pi\tau}. \label{Bshear}
\end{align}
The term with coefficient $\chi$ in the above equation contains 
correction only due to third-order. The first-order shear 
expression, $\pi=4\beta_\pi\tau_\pi/3\tau$, has been used to 
rewrite some of the third-order contributions in the form 
$\pi^2/(\beta_\pi\tau)$. The transport coefficients in our 
calculation simplify to
\begin{equation}\label{BTC}
\tau_\pi = \frac{\eta}{\beta_\pi}, \quad \beta_\pi = \frac{4P}{5}, \quad \lambda = \frac{38}{21}, \quad \chi = \frac{72}{245}.
\end{equation}
We compare these transport coefficients with those obtained from 
Eq. (\ref{TOEF}), where they reduce to
\begin{equation}\label{BTCE}
\tau_\pi' = \frac{\eta}{\beta_\pi'}, \quad \beta_\pi' = \frac{2P}{3}, \quad \lambda' = \frac{4}{3}, \quad \chi' = \frac{3}{4}.
\end{equation}

For comparison, we also state the exact solution of Eq. (\ref {RBE}) 
in one-dimensional scaling expansion \cite{Baym:1984np,Florkowski:2013lza}:
\begin{equation}\label{ESBE}
f(\tau) = D(\tau,\tau_0)f_{\rm in} + \int_{\tau_0}^{\tau}\frac{d\tau'}{\tau_R(\tau')}D(\tau,\tau')f_0(\tau'),
\end{equation}
where, $f_{\rm in}$ and $\tau_0$ is the initial distribution 
function and proper time respectively, and
\begin{equation}\label{DT2T1}
D(\tau_2,\tau_1) = \exp\left[-\int_{\tau_1}^{\tau_2}\frac{d\tau''}{\tau_R(\tau'')}\right].
\end{equation}
The damping function, $D(\tau_2,\tau_1)$, has the properties 
$D(\tau,\tau)=1$, $D(\tau_3,\tau_2)D(\tau_2,\tau_1)=D(\tau_3,\tau_1)$, 
and
\begin{equation}\label{DREL}
\frac{\partial D(\tau_2,\tau_1)}{\partial\tau_2} = -\frac{D(\tau_2,\tau_1)}{\tau_R(\tau_2)}.
\end{equation}
To obtain the exact solution, the Boltzmann relaxation time is taken 
to be the same as the shear relaxation time $(\tau_R=\tau_\pi)$. The 
hydrodynamic quantities can then be calculated by using Eq. (\ref 
{ESBE}) for the distribution function in Eq. (\ref{NTD}) and 
performing the integrations numerically.

\begin{figure}[t]
\begin{center}
\includegraphics[scale=0.33]{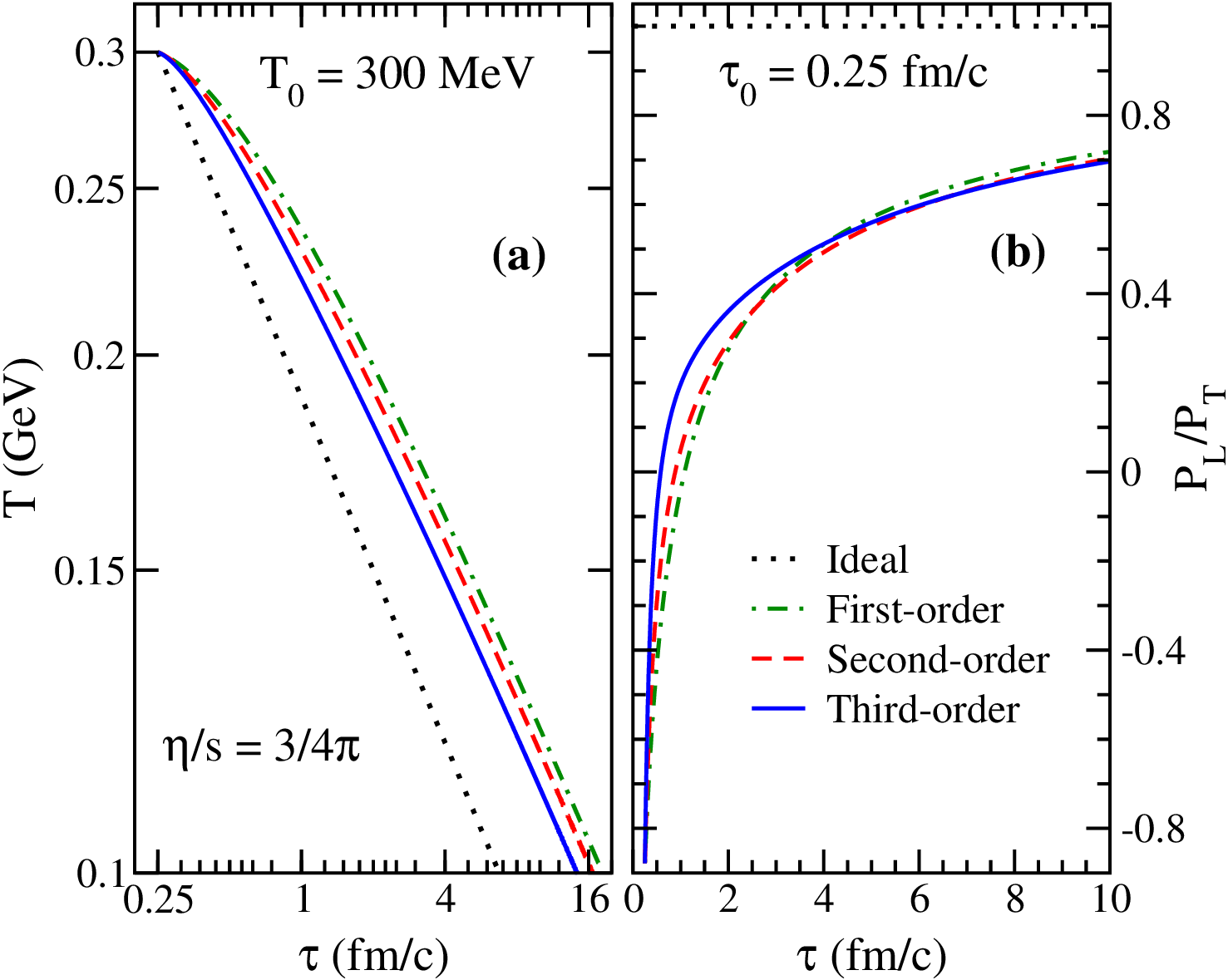}
\end{center}
\vspace{-0.4cm}
\caption{(Color online) Time evolution of (a) temperature and (b) 
  pressure anisotropy ($P_L/P_T$), in ideal (dotted line), 
  first-order (dashed-dotted lines), second-order (dashed line) and 
  third-order (solid lines) hydrodynamics, for Navier-Stokes initial 
  condition, $(\pi_0=4\eta/3\tau_0)$.}   
\label{TPLPT}
\end{figure}

To quantify the differences between ideal, first-order, 
second-order, and third-order theories, we solve the evolution 
equations with initial temperature $T_0=300$ MeV at initial time 
$\tau_0=0.25$ fm/c. These values correspond to the Relativistic 
Heavy-Ion Collider initial conditions \cite{El:2007vg}. Figure \ref 
{TPLPT} shows proper time evolution of temperature and pressure 
anisotropy $P_L/P_T\equiv(P-\pi)/(P+\pi/2)$ in ideal (dotted line), 
first-order (dashed-dotted lines), second-order (dashed line) and 
third-order (solid lines) hydrodynamics. Here we have assumed 
Navier-Stokes initial condition for shear pressure 
$(\pi_0=4\eta/3\tau_0)$ and solved the evolution equations for a 
representative shear viscosity to entropy density ratio, 
$\eta/s=3/4\pi$.

In Fig. \ref{TPLPT} (a), we observe that while ideal hydrodynamics 
predicts a rapid cooling of the system, evolution based on 
third-order equation also shows faster temperature drop compared to 
first-order and second-order evolutions. This implies that the 
thermal photon and dilepton spectra, which are sensitive to 
temperature evolution, may be suppressed by including third-order 
corrections. Moreover, with third-order evolution, the freeze-out 
temperature is attained at an earlier time which may affect the 
hadronic spectra as well. In Fig. \ref {TPLPT} (b), note that at 
early times the third-order evolution results in faster 
isotropization of pressure anisotropy compared to first-order and 
second-order. However at later time, the pressure anisotropy 
obtained using second and third-order equations merge indicating the 
convergence of gradient expansion in fluid dynamics.

\begin{figure}[t]
\begin{center}
\includegraphics[scale=0.4]{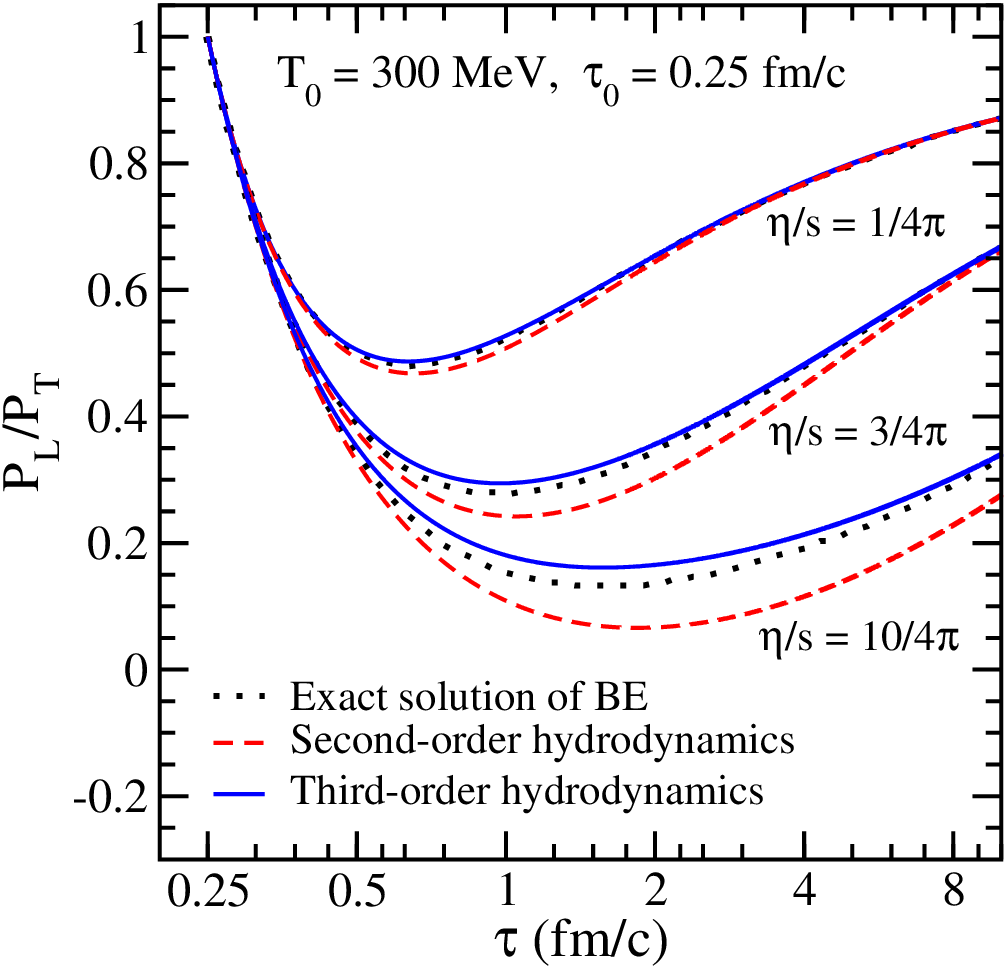}
\end{center}
\vspace{-0.4cm}
\caption{(Color online) Time evolution of $P_L/P_T$ obtained 
  using exact solution of Boltzmann equation (dotted line), 
  second-order equations (dashed lines), and third-order equations 
  (solid lines), for isotropic initial pressure configuration 
  $(\pi_0=0)$ and various $\eta/s$.} 
\label{PLPTR}
\end{figure}

Figure \ref{PLPTR}, shows the proper time dependence of pressure 
anisotropy for various $\eta/s$ values with isotropic initial 
pressure configuration, i.e., $\pi_0=0$. The improved agreement of 
third-order results (solid lines) with the exact solution of BE 
(dotted line) as compared to second-order results (dashed line) also 
suggests the convergence of the derivative expansion in hydrodynamics.

\begin{figure}[t]
\begin{center}
\includegraphics[scale=0.4]{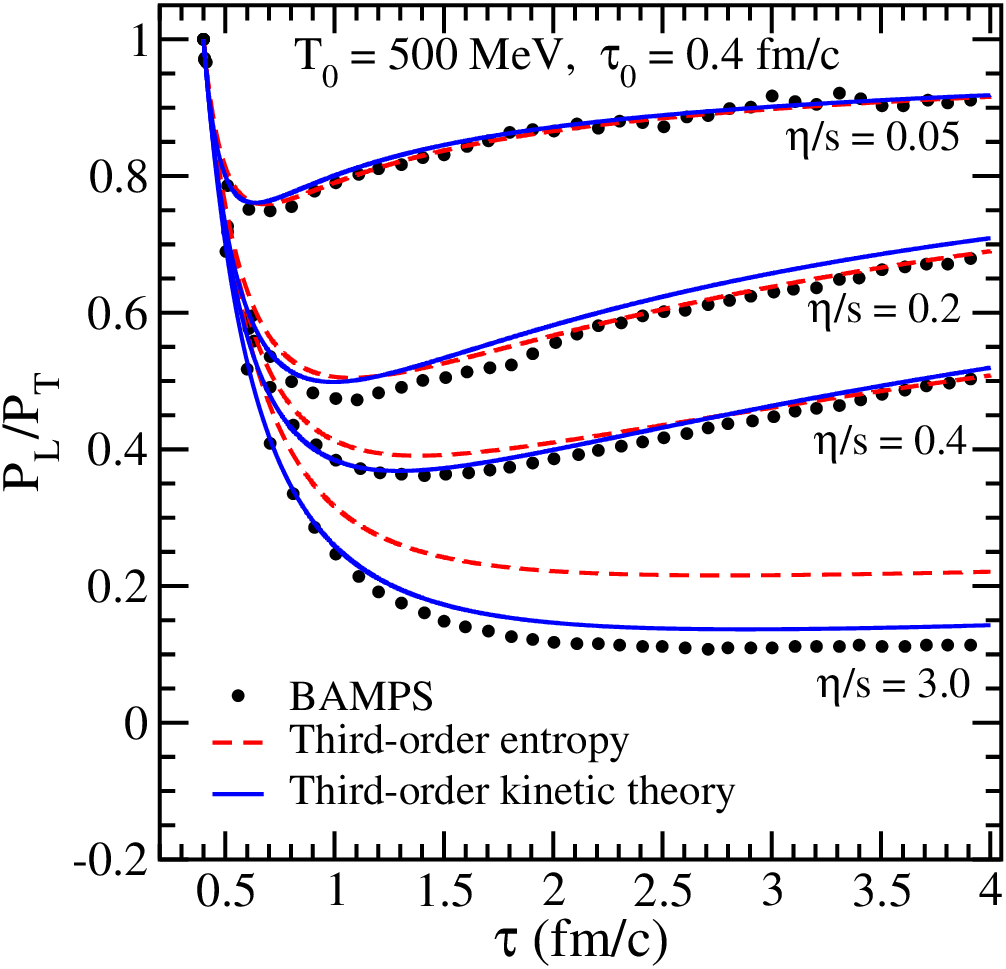}
\end{center}
\vspace{-0.4cm}
\caption{(Color online) Time evolution of $P_L/P_T$ in BAMPS (dots), 
  third-order calculation from entropy method, Eq. (\ref{TOEF})
  (dashed lines), and the present work (solid lines), 
  for isotropic initial pressure configuration $(\pi_0=0)$ and various $\eta/s$.} 
\label{PLPTL}
\end{figure}

Figure \ref{PLPTL}, also shows the time evolution of pressure 
anisotropy for initial temperature $T_0=500$ MeV at initial time 
$\tau_0=0.4$ fm/c which corresponds to Large Hadron Collider initial 
conditions \cite{El:2007vg}. The initial pressure configuration is 
assumed to be isotropic and the evolution is shown for various 
$\eta/s$ values. The solid lines represent the results obtained in 
the present work by solving Eqs. (\ref {BED}) and (\ref {Bshear}) 
with transport coefficients of Eq. (\ref {BTC}). The dashed lines 
corresponds to results of another third-order theory derived based 
on second-law of thermodynamics with transport coefficients given in 
Eq. (\ref {BTCE}). The dots represent the results of numerical 
solution of BE using a transport model, the parton cascade BAMPS 
\cite {El:2009vj,Xu:2004mz}. The calculations in BAMPS are performed 
by changing the cross section such that $\eta/s$ remains constant. 
While the results from entropy derivation overestimate the pressure 
anisotropy for $\eta/s>0.2$, those obtained in the present work 
(kinetic theory) are in better agreement with the BAMPS results. 

The RTA for the collision term in BE is based on the assumption that 
the effect of the collisions is to exponentially restore the 
distribution function to its local equilibrium value. Although the 
information about the microscopic interactions of the constituent 
particles is not retained here, it is a reasonably good 
approximation to describe a system which is close to local 
equilibrium. It is important to note that although the third-order 
viscous equations derived here uses BE with RTA for the collision 
term, the evolution shows good quantitative agreement with BAMPS 
results which employs realistic collision kernel \cite {Xu:2004mz}. 
Indeed in Ref. \cite{Dusling:2009df}, it has been shown that for a 
purely gluonic system at weak coupling and hadron gas with large 
momenta, BE in RTA is a fairly accurate description. Furthermore, 
the experimentally observed $1/\sqrt{m_T}$ scaling of the HBT radii, 
which was shown to be broken by including viscous corrections to the 
distribution function \cite {Teaney:2003kp}, can be restored by 
using the form of the non-equilibrium distribution function obtained 
here \cite {Bhalerao2}. All these factors clearly suggests that the 
BE in RTA can be applied quite successfully in understanding the 
hydrodynamic behavior of the strongly interacting matter formed in 
heavy-ion collisions.


To summarize, we have derived a novel third-order evolution equation 
for the shear stress tensor from kinetic theory within relaxation 
time approximation. Instead of Grad's 14-moment approximation, 
iterative solution of Boltzmann equation was used for the 
nonequilibrium distribution function and the evolution equation for 
shear tensor is derived directly from its definition. Within 
one-dimensional scaling expansion, we have demonstrated that the 
third-order hydrodynamics derived here provides a very good 
approximation to the exact solution of Boltzmann equation in 
relaxation time approximation. Our results also show a better 
agreement with the parton cascade BAMPS for the $P_L/P_T$ evolution 
compared to those obtained from entropy derivation.

\begin{acknowledgments}
The author thanks Rajeev S. Bhalerao and Subrata Pal for helpful 
discussions, and Jasmine Sethi for a critical reading of the 
manuscript.
\end{acknowledgments}

\end{document}